# Intersubband polaritonic metasurfaces for high-contrast ultra-fast power limiting and optical switching

Michele Cotrufo[1,2,†], Jonas Krakofsky[3,†], Sander Mann[1,†], Gerhard Böhm[3,*], Mikhail A. Belkin[3,*], Andrea Alù[1,4,*]

[1]Photonics Initiative, Advanced Science Research Center, City University of New York, New York, NY 10031, USA

[2]The Institute of Optics, University of Rochester, Rochester, New York 14627, USA

[3]Walter Schottky Institute, Technical University of Munich, 85748 Garching, Germany

[4]Physics Program, Graduate Center of the City University of New York, New York, NY 10016, USA

[†] These authors contributed equally

*aalu@gc.cuny.edu; mikhail.belkin@wsi.tum.de

## Abstract

Nonlinear intersubband polaritonic metasurfaces support one of the strongest known ultrafast nonlinear responses in the mid-infrared frequency range across all condensed matter systems. Beyond harmonic generation and frequency mixing, these nonlinearities can be leveraged for ultrafast optical switching and power limiting, based on tailored transitions from strong to weak polaritonic coupling. Here, we demonstrate synergistic optimization of materials and photonic nanostructures to achieve large reflection contrast in ultrafast polaritonic metasurface limiters. The devices are based on optimized semiconductor heterostructure materials that minimize the intersubband transition linewidth and reduce absorption in optically saturated nanoresonators, achieving a record-high reflection contrast of 54% experimentally. We also discuss opportunities to further boost the metrics of performance of this class of ultrafast limiters, showing that reflection contrast as high as 94% may be realistically achieved using all-dielectric intersubband polaritonic metasurfaces.

## Introduction

Nonlinear optical materials with large and fast Kerr ($\chi^{(3)}$) nonlinearity are highly desirable for a wide range of applications, including frequency combs and short laser pulse generation, protection of sensitive equipment via optical power limiting, photonic information processing, and all-optical computing [1]. To address these needs, a number of approaches have been pursued with the goal of achieving large values of $\chi^{(3)}$ in metasurfaces, thin-film and two-dimensions materials, such as using intrinsic metal or dielectric nonlinearities [2–6], nonlinearities of semiconductors and conduction-to-valence band transitions in semiconductor heterostructures [7,8], two-dimensional materials [9,10], epsilon-near-zero materials [11–15] and J-aggregates or other molecules [16–18]. Large nonlinearities and subwavelength optical thicknesses enable operation with very tight light focusing in the focal plane, while avoiding the phase-matching constraints of bulk nonlinear materials.

Sub-picosecond Kerr nonlinearities have been recently reported in intersubband polaritonic metasurfaces operating in in the mid-infrared ($\lambda \approx$ 3-30 μm) spectral range [19]. The operating principle of these structures (Figs. 1a-b) is based on a fast change from strong to weak coupling between a cavity mode and an intersubband (ISB) transition in a multi-quantum-well (MQW) semiconductor heterostructure [20–23]. These metasurfaces yield – to date – the largest picosecond third-order nonlinear response in condensed matter systems in the infrared spectral range, with effective $\chi^{(3)}$ values on the order of 3.4×10$^{-13}$ m$^2$/V$^2$, and a response time shorter than 2 ps for operating wavelengths around 7.5 µm [19].

While large nonlinearities are beneficial in reducing the required intensity levels, for practical applications such as optical power limiting, saturable absorber mirrors for laser mode-locking and frequency comb generation in the mid-infrared spectral range, and the development of nonlinear optical elements for optical computing and all-optical control, achieving large absorption/reflection contrast is equally important. Here and in the following, we define the reflection contrast of a metasurface limiter as the difference between the minimum and maximum reflection levels within a certain input power range. The metasurfaces reported in Ref. [19] and similar structures reported recently in [24] displayed moderately low power-dependent reflection contrast, with a maximum reflectivity change of about 25-30% reported in [19].

In this work, we present Kerr nonlinear metasurfaces based on intersubband polaritons which achieve a nonlinear absolute reflection contrast of 54%, while simultaneously displaying record-high values of effective $\chi^{(3)}$. Such enhancement is due to the combination of an optimized MQW system based on $Ga_{0.53}In_{0.47}As/GaAs_{0.51}Sb_{0.49}$ heterostructures, supporting ISB transitions with extremely narrow linewidths, more than a factor of 4 narrower than those of the $Al_{0.48}In_{0.52}As/Ga_{0.53}In_{0.47}As$ heterostructures used in Ref. [19], optimized nanophotonic structures leveraging these material resonances, and vertical, rather than diagonal, intersubband transitions [19], which maximize the transition dipole moment. The measured reflection contrast of 54% is, in fact, limited by the range of input powers available in our setup, and we estimate that reflection contrasts above 60% can be observed within the same realized device. Besides largely increasing the reflection contrast, the narrow linewidth of our ISB transitions also unveils exotic absorption features associated to the bending of the semiconductor bands close to the Schottky metal contacts, which may be leveraged to further improve metasurface performance. The main factor limiting the reflection contrast performance in our devices is the residual absorption in the metallic nanoresonators. We theoretically show that, by combining our optimized MQW system with all-dielectric metasurfaces, reflection contrasts as high as 94% may be realistically achieved.

## Results

**Physical model.** We begin by briefly outlining the physics behind the nonlinear response of a polaritonic metasurface induced by the saturation of ISB transitions in MQWs [19]. We consider a metasurface with a unit cell schematically shown in Fig. 1a, formed by a stack of MQW layers interacting with a metal patch antenna — a platform that has been investigated in several recent works [19,23,24]. This system is analyzed by a combined coupled-mode theory/Maxwell-Bloch (CMT/MB) model [19], modified to account for the absorption from higher-energy electron subbands, and described by the equations

$$\frac{d}{dt}a = [i\omega_a - \gamma_r - \gamma_a - \gamma_s(w+1)]a + iNgq + \frac{\sqrt{2\gamma_r}}{\hbar\omega A}s_+, \qquad (1)$$

$$\frac{d}{dt}q = (i\omega_q - \gamma_2)q - igwa, \qquad (2)$$

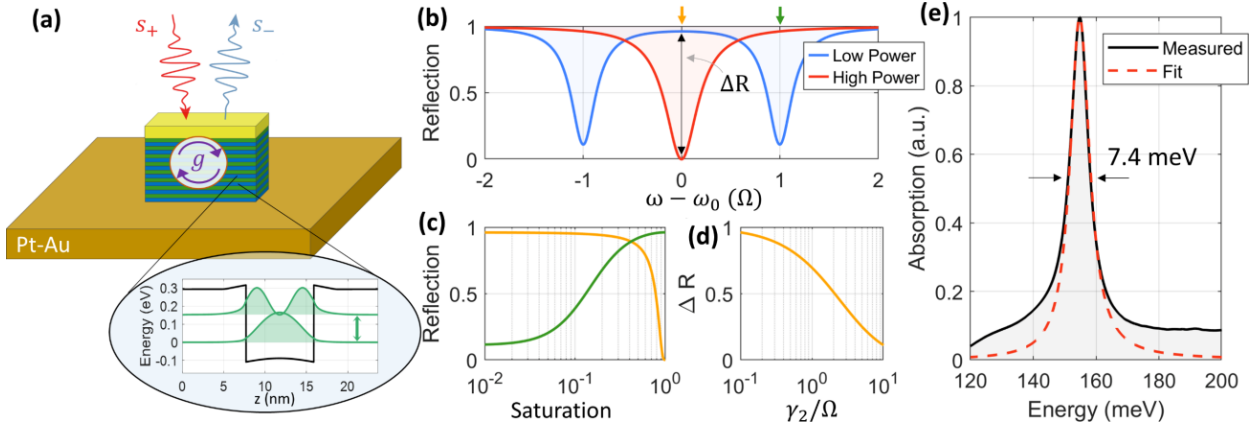

**Figure 1. (a)** Schematic of the polaritonic metasurface. An incoming field ($s_+$) feeds a nanoantenna, which in turn couples to the ISB transitions of the MWQ stack with a Rabi rate $g$, and determines the outgoing field ($s_-$). **(b)** In the strong coupling regime, the low-power reflection spectrum of the system (blue line) displays the characteristic polaritonic splitting, leading to very large reflection at $\omega = \omega_0$. For high impinging intensities (red line) the ISB transitions in all quantum wells are fully saturated, and the reflection spectrum approaches the one of the bare nanoantenna, featuring very low reflection at $\omega = \omega_0$. In these calculations we assumed $\gamma_a = \gamma_r = \gamma_2 = \Omega/10$. **(c)** Reflection spectra of the metasurface versus the saturation level of the MQW for two different impinging frequencies: $\omega = \omega_0$ (orange line in panel (c), also marked in panel (b) with the orange arrow) and $\omega = \omega_0 + \Omega$ (green line in panel (c), also marked in panel (b) with the green arrow). **(d)** Reflection contrast $\Delta R$ defined as the difference in the low- and high-intensity reflection coefficients of the metasurface versus $\gamma_2/\Omega$, for $\omega = \omega_0$. In this plot, the value of $\gamma_2$ is varied while all other parameters are the same as in panel b. **(e)** Measured absorption spectra (black line) of the bare InGaAs/GaAsSb MQW material. By fitting the main peak with an analytical model (red dashed line) we extract a transition energy of about 153 meV and a linewidth equal to FWHM = $2\hbar\gamma_2$ = 7.4 meV.

$$\frac{d}{dt}w = 4\text{Im}(qa^*) + \gamma_1(w + 1), \quad (3)$$

$$s_- = s_+ - \sqrt{2\gamma_r}(\hbar\omega A)a. \quad (4)$$

The patch antenna is modeled as a low-loss optical resonator with resonant frequency $\omega_a = 2\pi\nu_a$, complex modal amplitude $a$ (normalized so that $|a|^2$ is the number of stored photons), and radiative and nonradiative decay rates $\gamma_r$ and $\gamma_a$, respectively. The optical resonance of the antenna can be directly excited by an incident plane wave with time-dependent amplitude $s_+$ (normalized so that $|s_+|^2$ is the incident power per unit cell area $A$) and carrier frequency $\omega$. The back-propagating outgoing field $s_-$ (Eq. 4) is given by the superposition of the direct reflection of the incident field and the field leaked by the optical resonance. The MQW is modeled by $N$ identical two-level quantum oscillators, each with transition frequency $\omega_q = 2\pi\nu_q$, total decoherence rate $\gamma_2$ and relaxation rate $\gamma_1$, coupled to the optical resonance with rate $g = \frac{d_{12}}{\hbar}\sqrt{\frac{\hbar\omega}{2\varepsilon_0\varepsilon_r V}}$, where $\varepsilon_r$ is the relative permittivity of the MQW system at zero doping, $d_{12}$ is the transition dipole moment from state 1 to state 2, and V is the mode volume. This effective two-level system describes the ISB electronic transition between the ground state and the first excited state of the conduction band (inset of Fig. 1a), assuming that $N$ electrons are in the ground state at thermal equilibrium. The state of each two-level emitter is described by the off-diagonal element of the reduced density matrix, denoted by $q$ (Eq. 2), which describes the polarization of the transition, and by its inversion $w = n_e - n_g$, where $n_g$ and $n_e$ are the electron populations in the ground and first excited electron levels in the MQW system, normalized to the total electron population. Additionally, Eq. (1) contains an effective loss term $\gamma_s(w + 1)$ [19], which accounts for light absorption due to transitions of the excited electrons in higher energy states. This term becomes relevant at high pump intensities, depending on how significant this additional absorption is compared to the nonradiative decay rate $\gamma_a$ of the cavity and the Rabi splitting (see Eq. (5)) at low intensity.

The inversion $w$ is the source of nonlinearity in our system. At low intensities ($|a| \approx 0$), the inversion approaches $w \to -1$, and Eqs. (1)-(2) describe two classical coupled oscillators. In this linear regime, when the system is degenerate ($\omega_0 = \omega_a = \omega_q$) and in strong coupling, the reflection spectrum $R(\omega) \equiv |s_-(\omega)|^2/|s_+(\omega)|^2$ (blue line in Fig. 1b) features a characteristic Rabi splitting, displaying two distinct polaritonic resonant dips with equal linewidths and depth, and with resonant frequencies symmetrically detuned from the central frequency $\omega_0$ by $\pm\Omega$, where $\Omega \equiv \sqrt{N}g$. As a result of this splitting, the reflection is large at $\omega = \omega_0$, approaching the limit $R(\omega_0) \to 1$ for $\Omega \gg \{\gamma_1, \gamma_2, \gamma_a, \gamma_r, \gamma_s\}$. As the impinging intensity increases, the saturation of the ISB transition causes the inversion to approach zero ($w \to 0$), and the ISB transitions (Eq. 2) are completely decoupled from the optical resonance. In this regime, the reflection spectrum of the system approaches the one of the bare patch antenna (red line in Fig. 1b), consisting of a single reflection dip centered at the frequency $\omega = \omega_0 = \omega_a$. Because of this saturation effect, the energy stored in the optical resonance, and thus the outgoing field (Eq. 4), depends nonlinearly on the incident intensity $|s_+|^2$. In particular, at a fixed excitation frequency $\omega = \omega_0$ (yellow arrow in Fig. 1b) the reflection strongly decreases as the input intensity (and thus the emitter inversion) increases. Instead, for excitation frequencies close to the frequencies of the polaritonic dips ($\omega_0 \pm \Omega$, green arrow in Fig. 1b), the reflection level evolves from low (at low intensities) to high values (at high intensities). The evolution of the reflection versus saturation, defined as $w + 1$, is displayed in Fig. 1c for excitation frequencies equal to $\omega_0$ (yellow line) and $\omega_0 + \Omega$ (green line).

The relaxation time $1/\gamma_1$ is typically the slowest time constant in Eqs. (1)-(3). Thus, it dictates the timescales over which the inversion and saturation evolve in time, and it determines the speed of the metasurface nonlinear response. Values of $1/\gamma_1 \approx$ 1-2 ps have been determined experimentally [19], confirming that this system can be used as a fast power limiter or a fast saturable mirror. Another important figure of merit towards the practical application of these devices for power limiting is the reflection contrast $\Delta R$, defined as the difference between the reflection of the system in the fully linear case ($w = -1$) and the fully saturated case ($w = 0$), i.e., $\Delta R \equiv R(w = -1) - R(w = 0)$ (see also vertical black arrow in Fig. 1b). For an ideal power limiter $\Delta R = 1$, while an ideal saturable absorber features $\Delta R = -1$.

**Reflection contrast limits.** In [19] we showed that the achievable reflection contrast $\Delta R$ is fundamentally limited by the different decay channels of the system $\gamma_2$, $\gamma_a$, $\gamma_s$, and $\gamma_r$. In particular, if low reflection at high powers is desired, i.e., for a power limiter operation, the system needs to be close to critical coupling when fully saturated, i.e., $\gamma_r \approx \gamma_a + \gamma_s$. In this limit, we find that

$$|\Delta R| = \frac{\left(1 - \frac{\gamma_s \gamma_2}{\Omega^2}\right)^2}{\left(1 + 2\frac{\gamma_a \gamma_2}{\Omega^2} + \frac{\gamma_s \gamma_2}{\Omega^2}\right)^2} \leq \frac{1}{\left(1 + 2\frac{\gamma_a \gamma_2}{\Omega^2}\right)^2}, \tag{5}$$

where the upper limit of inequality is achieved when $\gamma_s = 0$. Equation (5) shows that $\Delta R$ is limited by the largest decay channel of the system. Since the ratio $\gamma_2/\Omega$ is present in all the 'damping' terms in Eq. (5), large values of $|\Delta R|$ can be achieved with ISB systems featuring small decoherence rates $\gamma_2$ and large vacuum Rabi frequencies $\Omega$, while keeping $\gamma_a$ and $\gamma_s$ as small as possible [19]. As an example, Fig. 1d shows the calculated $\Delta R$ as a function of $\gamma_2/\Omega$ and for a fixed value of $\gamma_s = 0$ and $\gamma_a/\Omega = 0.1$.

In our earlier work [19], values of $\gamma_a/\Omega \approx 0.04$, $\gamma_2/\Omega \approx 0.87$ and $\gamma_s/\Omega \approx 0.37$ were demonstrated. Based on Eq. (5), for these parameters the maximum theoretical value of the reflection contrast is $|\Delta R| \approx 0.23$ or 23%, which matches well the response reported in [19]. As follows from Eq. (5), we can improve the reflection contrast by increasing the value of $\Omega$, reducing the product of $\gamma_a$ and $\gamma_2$, and minimizing $\gamma_s$. We analyze the practical limits of each of these approaches in the following. For a given operating frequency ω, the vacuum Rabi frequency can be written as [25]

$$\Omega = \sqrt{\frac{e^2}{4\varepsilon_0 \varepsilon_r m_e^*} \frac{f_{12}(N_1 - N_2)}{h_{QW}}}, \qquad (6)$$

where $e$ is the electron charge, $m_e^*$ is the effective electron mass, $f_{12} = 2m_e^*\omega_q d_{12}^2/\hbar$ is the oscillator strength of the intersubband transition coupled to the cavity mode, $N_1$ and $N_2$ are the 2D electron densities in the ground and excited states in a single MQW period, and $h_{QW}$ is the thickness of a single MQW period. The maximum oscillator strength $f_{12}$ is achieved for a vertical transition in a quantum well, with the value $f_{12} \approx 0.96$ for an infinite quantum well [26]. Assuming a highly degenerate 2D electron gas, the maximum value of concentration difference is $(N_1-N_2) = \frac{m_e^*\omega_q}{\pi\hbar}$, which is achieved when the Fermi energy is $E_F = \hbar\omega_q$. For larger doping levels, both ground and excited states are being filled with electrons at the same rate. The minimum MQW period $L$ is set by the requirement that a single quantum well supports a transition at the desired energy $\hbar\omega_q$, which determines the minimum quantum well width to be $h_{QW} = \sqrt{\frac{3\pi^2\hbar}{2m_e^*\omega_q}}$. In practice, the value of $L$ is always larger than $h_{QW}$ because a quantum barrier is also required to separate individual quantum wells (in the limit of high conduction band offset, the barrier width can be very small, but in realistic systems it is often comparable to $L_{QW}$). By including these equalities into Eq. (6), we find that the maximum value of vacuum Rabi frequency is

$$\Omega_{\max} = \omega_q \sqrt{\frac{e^2 f_{12}}{2\pi^2 \hbar \varepsilon_0 \varepsilon_r} \sqrt{\frac{2m_e^*}{3\hbar\omega_q}}}. \qquad (7)$$

This value shows a weak dependence on the effective electron mass, and thus the choice of effective mass in the MQW system is not of particular importance [27]. The relative dielectric constant can be assumed to be $\varepsilon_r \approx 10$ for most suitable ISB systems [27]. Assuming $f_{12} = 0.96$, $m_e^* \approx 0.04 m_e$, where $m_e$ is the mass of a free electron, and $\hbar\omega_q \approx 150$ meV we obtain $\Omega_{\max} \approx 0.26\,\omega_q$.

In the system reported in Ref. [19] we obtained $\Omega = 0.13\ \omega_q$, which is lower than $\Omega_{max}$ by a factor of ~2. This can be attributed to a combination of several factors: first, the doping was ~2 times lower than the one required to achieve the optimal value of $N_1$-$N_2$; second, the use of a diagonal transition in the structure in Ref. [18] led to smaller than optimal transition oscillator strength.; finally, the barriers in the MQW structure had roughly the same thickness as the quantum wells (i.e., $L \approx 2L_{QW}$). While the first two issues could be addressed and improved, this would lead to only to a modest increase in $\Omega$. Moreover, significantly reducing the barrier thickness is not a viable route, as thinner barriers result in an energy level anticrossing between adjacent quantum wells, which increases the intersubband transition linewidth. We further note that all damping constants in Eq. (5) scale, in first approximation, with the operating frequency, and thus changing the transition frequency $\omega_q$ is not expected to improve the metasurface performance.

Based on this discussion, the Rabi frequency $\Omega$ cannot be increased significantly without compromising other performance metrics. Hence, we turn our attention to the possibility of increasing the reflection contrast $\Delta R$ by instead reducing the damping constants $\gamma_2$, $\gamma_a$ and $\gamma_s$ in Eq. (5). The value of $\gamma_a$ is related to the optical loss of the bare optical cavity, and it is due to Ohmic loss in the metal and scattering on the nanoresonator walls. In the improved metasurface design reported in the next section, we reduce the roughness of the nanoresonator sidewall via optimization of the fabrication process, while keeping the same nanoresonator design as in Ref. [19]. We note that the value of $\gamma_a$ can be also reduced by switching to dielectric nanoresonators [28,29], albeit at the expense of a somewhat reduced electromagnetic field confinement.

The value of $\gamma_s$ can be reduced by limiting the impact of the ISB transitions from the excited state to higher energy states. This approach is ultimately limited by the conduction band offsets available in mature MQW systems. The value of $\gamma_s$ for the maximum Rabi frequency can be estimated by assuming that all electrons are excited to a continuum of conduction band states, where they can be described as a free electron gas characterized by the Drude permittivity

$$\varepsilon(\omega) = \varepsilon_r - \frac{Ne^2}{\varepsilon_0 m_e^*(\omega^2 + i\omega/\tau)}. \tag{8}$$

We can use perturbation theory to find an approximate expression for the loss rate [30], leading to $\gamma_s \approx \omega_q \text{Im}(\varepsilon(\omega_q))/(2\varepsilon_r)$. Assuming the optimal bulk doping concentration of electrons $N = \frac{m_e^* \omega_q}{\pi \hbar h_{QW}} = 1.8 \times 10^{18}$ cm$^{-3}$ derived for Eq. (7), a typical electron scattering rate for heavily doped semiconductor materials $\tau \approx 0.1$ ps [31], and operating frequency $\hbar\omega = \hbar\omega_q \approx 150$ meV, we obtain an upper limit of $\gamma_s = 0.01\ \omega_q$. In our earlier work [19], we estimated $\gamma_s \approx 0.05\ \omega_q$ based on the nonlinear behavior of the polaritonic metasurface. However, we note that in Ref. [19] the quantum barriers in the MQW structure had a large conduction band offset (520 meV) and most of the electrons were not excited into the continuum for the range of intensities (up to 700 kW/cm$^2$) employed in the experiment. As discussed in [19], the value of $\gamma_s$ in this case instead likely captured intersubband absorption between higher bound states in the MQW system. In contrast, in the MQW system described in the next section the conduction band offset for the barriers is only 360 meV, facilitating electron excitation to the continuum via two-photon absorption processes, as in Fig. 1(a). Given a factor of 10 larger excitation intensities (up to 10 MW/cm$^2$) employed in the experiments reported here, Eq. (8) provides a good estimate for the upper limit of $\gamma_s$ for ideal doping levels. In practice, however, we operate in an intermediate intensity regime where the intersubband transition is nearly saturated, but the electrons are not fully ionized.

Finally, the ISB transition dephasing rate $\gamma_2$ is present in all the factors in Eq. (5). Thus, reducing the value of $\gamma_2$ can lead to significant improvements in the metasurface performance. The value of $\gamma_2$ is typically more than a factor of 10 larger than $\gamma_1$, and it is due to roughness-induced scattering at the quantum well/quantum barrier interfaces and electron-electron and phonon scattering [32]. For higher-doped materials, non-parabolicity [33] also contributes to increased $\gamma_2$.

**Record-contrast $\chi^{(3)}$ intersubband polaritonic metasurface.** It was recently reported that $Ga_{0.53}In_{0.47}As/GaAs_{0.51}Sb_{0.49}$ material systems [34] can produce ISB transitions with extremely narrow transition linewidths. The mechanism behind this material response has not yet been fully explained, but reduced interface roughness scattering and relatively smaller non-parabolicity likely plays a major role. Guided by the analysis in the previous section, we optimized the parameters of molecular beam epitaxy, such as growth speed and temperature, to significantly reduce the interface roughness scattering. As a result, we were able to produce $Ga_{0.53}In_{0.47}As/GaAs_{0.51}Sb_{0.49}$ heterostructures with similar doping levels as in the $Ga_{0.53}In_{0.47}As/Al_{0.48}In_{0.52}As$ system used in [19], but with a more than 4-fold reduction of the ISB transition linewidth.

Following the analysis presented above, this new MQW system was designed around a vertical ISB transition in a single quantum well, as shown in the inset of Fig. 1(a). Similar to the MQW systems used in previous metasurfaces, modulation doping was used to avoid scattering of electrons by impurities. The MQW stack was constructed by 15 repetitions, each consisting of an 8.2 nm wide $Ga_{0.53}In_{0.47}As$ well separated by a 15 nm wide $Ga_{0.53}In_{0.47}As$ barrier. Charge carriers within each well were introduced by two delta-like doping regions, with a concentration of $n = 6.5 \cdot 10^{11} \text{cm}^{-2}$, positioned 2.3 nm from each side of the well. To minimize band bending effects at the metal contacts, we have incorporated $Ga_{0.53}In_{0.47}As$ buffer layers with an $n$-doping of $1 \cdot 10^{19} \text{cm}^{-3}$ on both sides of the MQW stack. The bottom buffer layer has a 5 nm thickness, while the top layer also acts as an etch stop layer during fabrication and measures approximately 15 nm in thickness. This approach closely aligns with the design detailed in the previous work from Ref. [19]. Additional details of the sample wafer structure are given in Methods. The extremely narrow ISB absorption linewidth is confirmed via 45°-angled absorption measurements [33] shown in Fig. 1(e). The absorption measurement yields an ISB transition linewidth of $2\gamma_2 = 7.4$ meV, which is over a factor of 4 smaller than that of the MQW structures used in [19]. The measured spectrum yields an ISB transition energy of $\hbar\omega_q \approx 155$ meV, close to the simulated value. The height of the absorption peak can be used to deduce the actual doping level in our structure, which was found to be $N_b = 5.4 \times 10^{17}$ cm$^{-3}$. The corresponding sheet electron concentration density $n_{2D} = 1.3 \times 10^{12}$ cm$^{-2}$ is 2.2 times smaller than the optimal value of $2.5 \times 10^{12}$ cm$^{-2}$ derived in the previous section. The smaller-than optimal value of electron concentration originates from the nonlinear activation rates of silicon dopants experimentally observed in our material system. In particular, we observe that, in order to obtain the desired further two-fold increase of electron concentration under optimized growth conditions, it would be necessary to increase the concentration of silicon dopants by a factor of approximately 9. This much larger concentration of donor impurities increases the electron-impurity scattering rate, leading to a significant increase of the ISB transition linewidth, $\gamma_2$.

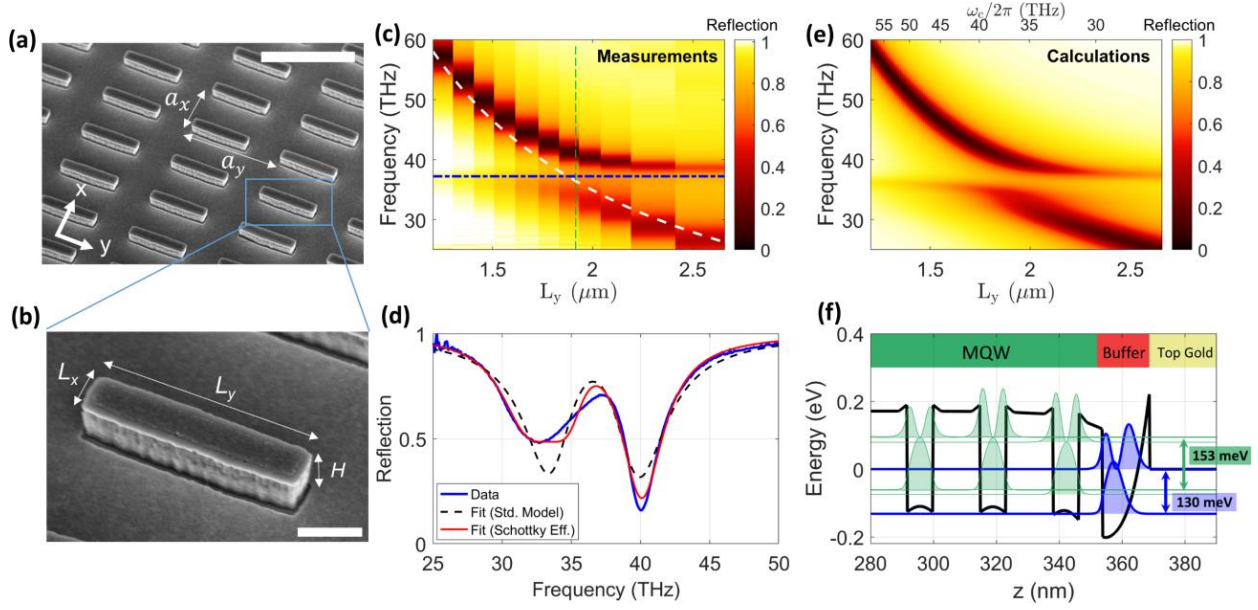

**Figure 2.** (a) SEM image of the metasurface, scalebar = 3 µm. (b) Zoomed-in SEM image, showing a single unit cell, scalebar = 500 nm. The lattice pitches are $a_x = 2$ µm and $a_y = 4$ µm and the antenna's dimensions are $L_x \approx$ 330 nm, $L_y \approx 1845$ nm and the MQW stack thickness is $H \approx 350$ nm. (c) Measured y-polarized reflection spectra of several devices as a function of $L_y$ (all other dimensions are as in panels a-b). The horizontal dashed-dotted blue line indicates the frequency of the bare ISB transition, while the dashed white line indicates the resonant frequency of the bare antenna. (d) Measured reflection spectra of the device with $L_y \approx 1845$ nm (blue solid line), corresponding to vertical dashed line in panel c. The other lines show the low-power reflection spectra calculated either without (black dashed line) or with (red line) the Schottky effect. (e) Calculated low-power reflection spectra of the metasurface, for the same geometries and frequency as in panel c, including the band structure bending caused by the Schottky effect. (f) Numerically calculated band structure of the MQW, including the Schottky effect at the top MQW-titanium/gold interface.

To demonstrate the improved performance unlocked by this MQW system, we used this material to fabricate and test polaritonic metasurfaces. Figures 2(a-b) show SEM pictures of a fabricated sample. The unit cell of the metasurface consists of a single patch antenna, obtained by sandwiching a stack of InGaAs/GaAsSb MQWs of thickness H≈350 nm between a flat layer of metal and a 50-nm-thick rectangular gold antenna with in-plane dimensions $L_x$ and $L_y$ (Fig. 2b). The patch antennas are arranged in periodic rectangular arrays with total lateral dimensions of approximately 200 µm (details on the fabrications are provided in the Supplementary Material). Because of the strong field confinement and the subwavelength unit cells, the optical response of the metasurfaces is governed by the resonance supported by each unit cell, which in turn is largely controlled by the antenna dimensions. In order to systematically tune the interaction between the ISB transition of the MWQ and the antennas, we fabricated devices with different antenna lengths in the range $L_y$ = 1 µm − 2.3 µm, while keeping lattice pitch ($a_x = 2$ µm and $a_y = 4$ µm) and the short side of the antenna ($L_x \approx 330$ nm) fixed.

The color-coded plot in Fig. 2c shows *y*-polarized reflection spectra of devices with different values of $L_y$, measured with a commercial Fourier-transform infrared spectroscopy (FTIR) system (see Methods). The measurements show the typical signature of strong coupling: two distinct reflection dips, corresponding to the MQW and antenna resonances, which anti-cross as the antenna resonance frequency (dashed white line) approaches the ISB frequency (horizontal dashed-dotted blue line). When the two bare resonance frequencies are approximately equal ($\nu_q = \nu_a \approx 38$ THz, corresponding to the vertical dashed green line in Fig. 2c, $L_y \approx 1845$ nm), the measured reflection spectrum (shown in Fig. 2d as solid blue line) features two reflection dips located at frequencies larger and smaller than the common frequency $\nu_0 = \nu_q = \nu_a$, and a reflection maximum at a frequency close to $\nu_0$. While the presence of these two polaritonic dips is expected in the strong coupling regime, the measured spectrum (solid blue lines in Fig. 2d) shows clear deviations from the conventional two resonator model (Eqs. 1-4 and Fig. 1b). First, the two polaritonic dips are not symmetrically detuned from the common resonant frequency $\nu_0$. Second, the two dips have different linewidths and depths, with the high-frequency dip being sharper and deeper than the low-frequency one. Such asymmetry is not possible based on the strong coupling of two oscillators (see also Fig. 1b), independently of their mutual coupling strength, decay rates, or other parameters, and it strongly contrasts with previous experimental results on similar devices [19]. The black dashed line in Fig. 2d shows the best fit of the experimental data by using Eqs. 1-4, emphasizing the incompatibility between the two-resonator model and the measured data.

These deviations from our theoretical model are due to the occurrence of an additional ISB transition localized at the top metal-semiconductor interface, where the Schottky effect, in combination with the n-doped top buffer layer, strongly affects the electronic band structure. We note that the effect of this additional transition becomes apparent only in the measurement of the patch antennas, and it does not appear in the absorption measurements of the bare MQW materials (Fig. 1e), since in that case no metal interface is located next to the buffer layer.

To corroborate this statement, we numerically calculated the band structure of the full MQW stack by using a self-consistent 8-band $\mathbf{k} \cdot \mathbf{p}$ model (Fig. 2f). In our model, we assumed that the top titanium/gold layer has a Schottky potential of 0.23 meV [35] and we included the effect of the band bending due to charge carrier separation at the Schottky interface. The band structure calculations display a clear bending of the conduction band potential energy (solid black line in Fig. 2f) near the metal interfaces and within the top buffer layer region (see the sample structure description in the Supplementary Material). As a consequence, the ISB transition near the interface has a transition energy (130 meV, blue arrow in Fig. 2f) lower than the one of the electronic states existing closer to the center of the MQW stack (153 meV, green arrow in Fig. 2f). The calculated charge carrier density for the barrier ISB transition is $N_b = 8 \times 10^{17} cm^{-3}$, which is 1.5 times larger than the charge carrier density inside the MQW stack ($N_c = 5.4 \times 10^{17} cm^{-3}$). We note that another ISB transition is expected to exist also at the bottom buffer layer. However, the bottom layer has a different geometry and smaller thickness (due to its different functionality), leading to a much larger ISB transition energy, exceeding 300 meV according to our calculations. Due to the much larger detuning, this additional ISB transition does not play a relevant role in our model.

In the presence of an ISB transition at approximately 130 meV in the top buffer layer, our metasurface spectra cannot be described by a standard two resonator model for the MQW-cavity interaction [19] [Eqs. (1)-(3)]. We emphasize that, while the additional ISB transitions are expected to exist in any polaritonic metasurface with MQWs close to metal layers [19,24], their effect is often hard to observe because of the large ISB transition linewidths. The effect instead is very evident here due to the very small linewidth of ISB transitions in our MQW material, which allows us to spectrally resolve these additional states. In order to validate and incorporate this effect into our model of the polaritonic metasurface, we expand our CMT/MB model by describing the dynamics of the MQW stack with two separate resonators, one describing the ISB transition at the center of the MQW (with electron density $N_c$) and the other one describing the ISB transition at the top buffer layer [see Fig. 2(f)] near the metal contact (with electron density $N_b$). The system is described by an expanded set of coupled equations:

$$\frac{d}{dt}a = (i\omega_a - \gamma_r - \gamma_a)a + iN_c g_c q_c + iN_b g_b q_b + \frac{\sqrt{2\gamma_r}}{\hbar\omega A}s_+, \tag{9}$$

$$\frac{d}{dt}q_c = (i\omega_{qc} - \gamma_{2c})q_c - ig_c w_c a, \tag{10}$$

$$\frac{d}{dt}q_b = (i\omega_{qb} - \gamma_{2b})q_b - ig_b w_b a, \tag{11}$$

$$\frac{d}{dt}w_c = 4\text{Im}(q_c a^*) + \gamma_1(w_c + 1), \tag{12}$$

$$\frac{d}{dt}w_b = 4\text{Im}(q_b a^*) + \gamma_1(w_b + 1), \tag{13}$$

where now $q_i$ and $w_i$ ($i = c,b$) are the polarizations and the inversions of the ISB transitions in the center *(c)* and in the buffer *(b)* layer, respectively. Each MQW transition has a different resonant frequency $\omega_{qi}$ and total decoherence rate $\gamma_{2i} = 1/T_{2i}$, while the relaxation rate $\gamma_1 = 1/T_1$ is assumed to be the same for simplicity. Moreover, each ISB transition couples to the cavity mode with a different Rabi frequency $\Omega_{Ri} = \sqrt{N_i}g_i$.

We begin by considering the low-power regime ($|a|^2 \approx 0$) where the inversions are $w_i = -1$, and we use this corrected model [Eqs. (9)-(13)] to fit the measured reflection spectra (Fig. 2c) of devices with different antenna lengths. In the fitting procedure we assume that $\omega_{qi}, \gamma_{2i}, \gamma_1$ and $\Omega_{Ri}$ are fixed across all devices (since all devices were made of the same MQW material) while the antenna parameters ($\omega_a, \gamma_r$, and $\gamma_a$) were allowed to vary. For the MQW material parameters, we obtained the values $\omega_{qc}/2\pi = 36.53$ THz, $\omega_{qb}/2\pi = 33.37$ THz, $\gamma_{2c}/2\pi = 0.88$ THz, $\gamma_{2b}/2\pi = 3.15$ THz, $\Omega_{Rc}/2\pi = 2.83$ THz, $\Omega_{Rb}/2\pi = 2.66$ THz. We note that the value of $\gamma_{2c}$ is in excellent agreement with the value independently extracted from the fit of the bare MQW absorption spectra in Fig. 1e, $\gamma_2/2\pi = 0.89$ THz. For the device corresponding to the vertical dashed green line in Fig. 2c, the retrieved antenna parameters are $\omega_a/2\pi = 36.82$ THz, $\gamma_r/2\pi = 0.87$ THz, $\gamma_a/2\pi = 1.46$ THz. The value of $\gamma_1$ was computed to be 0.9 THz from the optical-phonon-limited electron lifetime in the first excited electron subband in $In_{0.53}Ga_{0.47}As$ quantum wells, similar to the one computed and used for the MQW system in Ref. [19]. With the parameters obtained from the fit, and assuming that the antenna resonance wavelength scales linearly with its length, we can accurately reproduce the linear reflection measurements in Fig. 2(c), as shown by the color plot in Fig. 2e. The accuracy of our model is further revealed in Fig. 2d, where we plot the calculated reflection spectra when the antenna resonance frequencies and the $q_c$ transition frequencies are equal (solid red line), alongside with the corresponding measured data (blue line). Despite some residual discrepancies, the spectra calculated with the model in Eqs. (9)-(13) (red line) correctly reproduces the asymmetric shapes of the two polaritonic peaks observed in the experiment (blue line). Moreover, the value of the transition energy $\hbar\omega_{qb} = 138$ meV obtained from the fit agrees reasonably well with the one expected from the band structure calculations near a metal contact (130 meV, Fig 2f).

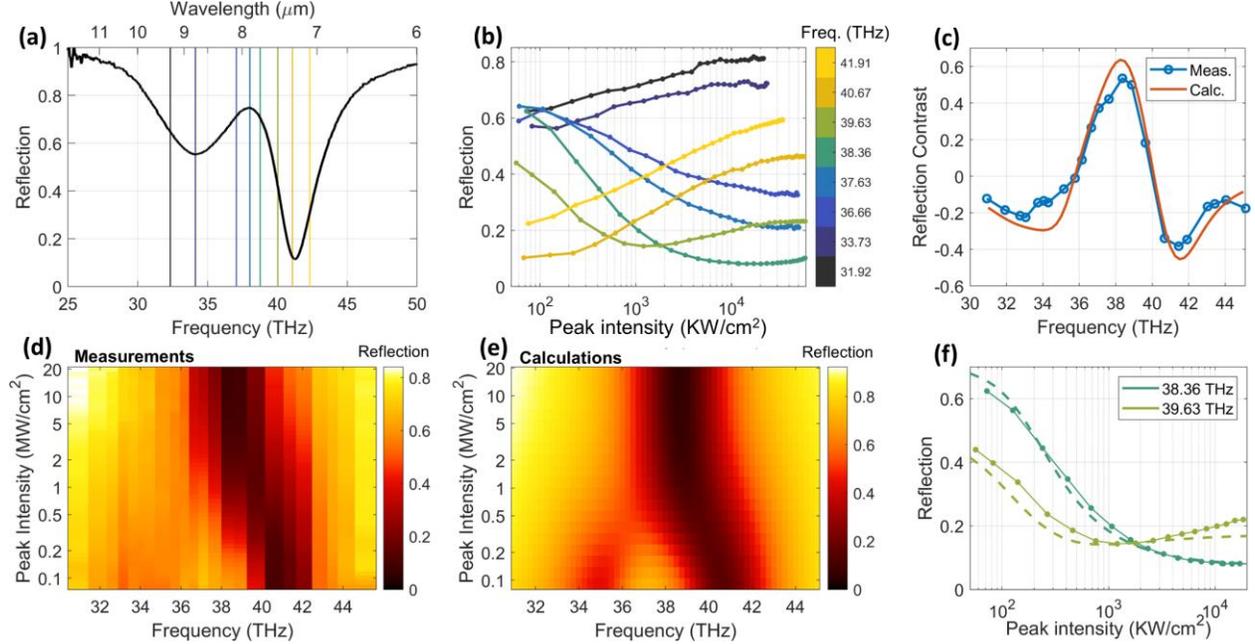

**Figure 3. Nonlinear response of the metasurface.** (a) Measured low-power reflection spectra of the device selected for the nonlinear measurements ($L_y \approx 1845$ nm), corresponding to the vertical dashed line in Fig. 2b. (b) Measured reflection versus impinging peak intensities at different selected frequencies. The excitation frequency in each measurement is color-coded (see colorbar) and indicated by a corresponding vertical line in panel a. (c) Experimentally measured (blue dots) and numerically calculated (orange line) reflection contrast versus impinging frequency. (d) Measured reflection versus impinging peak intensity and frequency, for a larger number of impinging frequencies. (e) Numerically calculated reflection, corresponding to the experimental data shown in panel d. (f) Comparison between measured (filled circles) and calculated (dashed lines) nonlinear reflection curves for two representative frequencies.

After having experimentally characterized and analytically interpreted the linear spectra of our polaritonic metasurfaces, we now show how their sharp spectral response, enabled by the narrow ISB linewidth, can be leveraged to achieve MQW-based saturable absorbers and power limiters with large reflection contrast. We focus on the device with spectrum shown in Fig. 2d (i.e. antennas with length $L_y \approx 1845$ nm), corresponding to the case where the ISB transition at the center of the MQW (described by $q_c$ in Eqs. 9-13) and the patch antenna are resonant. The corresponding reflection spectrum is reproduced in Fig. 3a. We experimentally measured the reflection of this metasurface versus impinging intensity for different frequencies across the metasurface spectrum. The measurements were performed with a custom-built setup described in the Supplementary Material. The excitation is provided by a pulsed laser with tunable carrier frequency, a repetition rate of 80 MHz and a pulse duration of 2 ps. Figure 3b shows the measured nonlinear reflection for 8 different frequencies, corresponding to the color-coded vertical lines in Fig. 3a. As expected from our theory [Figs. 1(b-c)], when the excitation frequency is close to the center of the gap between the two dips ($\nu = 38.36$ THz, dark green curve), the reflection strongly decreases with increasing intensity, starting from a large value of about 0.64 for intensities smaller than 100 KW/cm² and reaching values smaller than 0.1 for intensities of about 10 MW/cm², corresponding to a reflection contrast $|\Delta R| \approx 0.54$ or 54%.

For frequencies closer to the center of the dips the reflection evolves from low to high values as the intensity increases. In particular, when the excitation frequency is resonant with the narrowest dip ($\nu = 40.67$ THz, dark orange line in Figs. 3a and 3b), the reflection evolves from 0.1 to almost 0.5. Different nonlinear behaviors are obtained for other frequencies, including scenarios where the reflection level features a local minimum for intermediate intensity values ($\nu = 39.63$ THz, light green lines). We also note that, for most of the curves in Figs. 3b, the measured reflection contrast $\Delta R$ is limited by the intensity range available in our experiment. In particular, at the lowest intensity considered in our experiment (~ 50 KW/cm$^2$) the saturation level of the ISB transition appears to be already non-negligible, as manifested by the fact that several curves in Fig. 3b have a non-zero slope already at low intensities. Thus, as discussed later, the reflection contrast achievable with this metasurface is expected to be even larger than what observed in Fig. 3b. The color-coded plot in Fig. 3d shows the measured nonlinear reflection for a larger and more refined set of excitation frequencies. As clear from this plot, the reflection spectrum evolves from an asymmetric two-dip lineshape at low intensities to a single-dip lineshape at high intensities. The high-intensity spectrum corresponds to the case of a fully saturated ISB transition, and thus the single dip is attributed to the response of the bare antenna. The low reflection minimum achieved by the high-intensity spectra (min $[R_{\text{high}}(\nu)] \approx 0.1$) confirms that the antenna is close to the critical coupling condition $\gamma_a \approx \gamma_r$. From the measurements in Fig. 3d we can extract the frequency-dependent reflection contrast $\Delta R(\nu) \equiv R_{\text{low}}(\nu) - R_{\text{high}}(\nu)$, which is plotted in Fig. 3c (blue symbols). The experimental reflection contrast varies between $\Delta R \approx -0.4$ for the saturable absorber operation and $\Delta R \approx 0.54$ for the power limiter operation, which is approximately a factor of 2 larger than the previous state-of-the-art reported in [19].

   To model the nonlinear measurements in Fig. 3d, we numerically solved Eqs. (9)-(13) by assuming that the input $s_+(t)$ is a Gaussian pulse with duration of 2 ps, consistent with the experimental setting, and we calculated the resulting output field $s_-(t)$ for different input peak intensities. Importantly, in these nonlinear calculations all system parameters are fixed by the fit done on the linear measurements, i.e., Figs. 2c and 2e, except for the electron density of each ISB transitions, $N_c$ and $N_b$, which determine the corresponding saturation intensities. These values, together with the estimated volume of the unaffected and bent quantum wells [see Fig. 2(f)], yield $N_c = 5.4 \times 10^{17}$ cm$^{-3}$ and $N_b = 8 \times 10^{17}$ cm$^{-3}$. The numerical calculations (Fig. 3e) show excellent agreement with the experimental measurements, reproducing both the asymmetric lineshape at low intensities and the intensity-dependent evolution of the spectrum. The excellent agreement between simulations and measurements is highlighted in Fig. 3f, where we show the measured and calculated intensity-dependent reflection for two representative frequencies. Moreover, the frequency-dependent reflection contrast $\Delta R(\nu)$ extracted from the calculations (orange solid line in Fig. 3c) agrees well with the one extracted from measurements (blue circles in Fig. 3c), and it reproduces both the absolute values of $\Delta R(\nu)$ and the asymmetric tails. Overall, the excellent match between the experimental measurements and the theoretical predictions for both linear (Figs. 2c and 2e) and nonlinear (Figs. 3(c-f)) scenarios confirms that the dynamics of these polaritonic metasurfaces can be quantitatively described by expanding the simple two-level model [19] to account for the presence of additional ISB transitions, induced by the Schottky barrier formed near the interface between the MQW and the metal layers. As mentioned above, the measured reflection contrast is limited by the range of input power levels used in the experiment, and it could be increased by starting from lower input powers in the measurements in Fig. 3b. Indeed, the calculation shown in Fig. 3c (orange line) confirms that the reflection contrast of our device could be as high as $\Delta R \approx 0.64$ if a larger range of powers is considered.

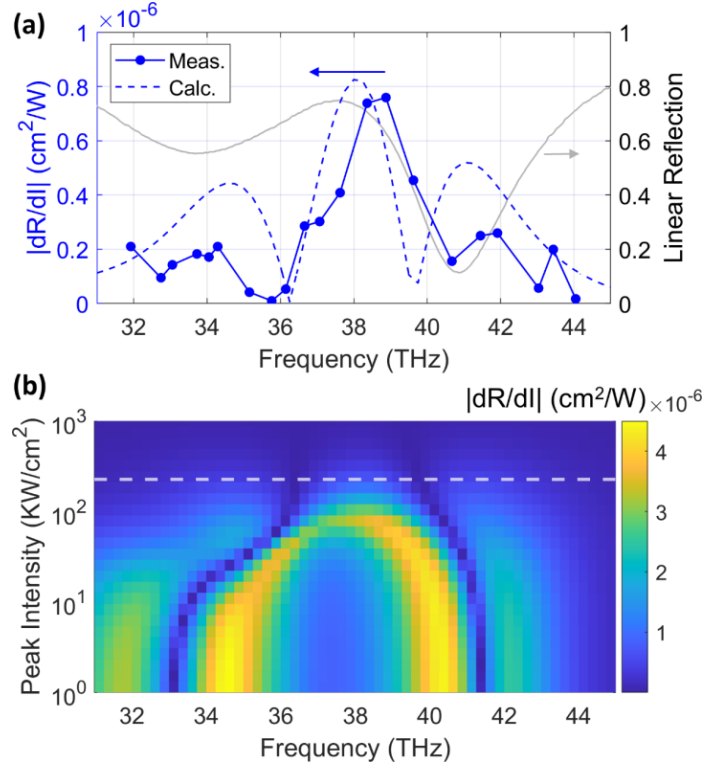

**Figure 4.** (a) Absolute values of $|dR/dI|$ extracted from measurements (filled blue circles) and calculations (dashed line) versus the impinging frequency, at a peak intensity of I ≈ 200 KW/cm² (horizontal dashed line in panel b). The linear spectrum of the device is also shown for reference (solid gray line, right vertical axis). (b) Calculated absolute values of the derivative $|dR/dI|$, versus impinging frequency and peak intensity. The horizontal dashed line denotes the intensity considered in panel a.

In the context of saturable materials and power limiting applications, an important figure of merit is the change of reflection (or absorption) upon small variations of the impinging intensity $I$, i.e., the derivative $dR/dI$. In Fig. 4a we show the experimental value of $|dR/dI|$ versus impinging frequency (blue symbols), extracted from the data in Fig. 3d at an impinging intensity of about 200 KW/cm². A slope $|dR/dI| \approx 0.8 \cdot 10^{-6}$ cm²/W is obtained for frequencies close to the center of the two-dip linear reflection spectra. The measured $|dR/dI|$ agrees well with values extracted from numerical simulations (dashed blue line in Fig. 4a). Since our system does not transmit any power due to the metal backing, the absorption is $A = 1 - R$, and thus $dA/dI = -dR/dI$. The slope $dA/dI$ can be used to estimate an effective nonlinear third-order imaginary susceptibility $\text{Im}[\chi^{(3)}_{\text{eff}}]$, via [19]

$$\text{Im}\left[\chi^{(3)}_{\text{eff}}\right] = \frac{2\varepsilon_0 n^2 c^2}{3\omega} \beta_{\text{eff}} \qquad (11)$$

where $\beta_{\text{eff}} \equiv \frac{dA}{dI}\frac{1}{2d}$ is the effective nonlinear absorption coefficient, and $d$ is the thickness of the metasurface. By assuming $n = 1$ and $d = 400$ nm, we obtain $\text{Im}\left[\chi^{(3)}_{\text{eff}}\right] = 2.2 \cdot 10^{-13}$ m²/V² for the central frequency $\nu_0 \approx 38$ THz. As mentioned above, the saturation of the ISB transitions is non-negligible already at the lowest available intensities in our experiment. This suggests that even larger values of $|dR/dI|$ may be observed at lower intensities. In order to numerically verify this effect, we repeated the calculations shown in Fig. 3e for lower impinging intensities (while keeping all the other parameters the same as in Fig. 3e), and we report the corresponding values of $|dR/dI|$ versus intensity and frequency in Fig. 4b. This calculation shows that derivatives as large as $|dR/dI| \approx 4.5 \cdot 10^{-6} \text{cm}^2/\text{W}$ are expected in our metasurface for peak intensities lower than 10 KW/cm$^2$. In turn, this value corresponds to an effective nonlinear susceptibility $\text{Im}\left[\chi_{\text{eff}}^{(3)}\right] > 10^{-12} \text{m}^2/\text{V}^2$, i.e., at least a factor of 3 times larger than the record nonlinearity reported in [19].

**Further improvements – dielectric intersubband polaritonic metasurfaces.** As discussed in Section 3, a key limiting factor in the polaritonic metasurfaces described above is the absorption rate $\gamma_a$, mostly due to the presence of metals. In principle, these absorption losses can be largely reduced by using all-dielectric metasurface designs [28,29]. Using the expressions derived for $\gamma_s$ and $\Omega$, assuming $\omega_q \gg 1/\tau$, and setting the absorption loss $\gamma_a$ to zero, we can rewrite Eq. 5 as

$$|\Delta R| \approx \frac{\left(1 - \frac{2\gamma_2}{\tau \omega_q^2 f_{12}}\right)^2}{\left(1 + \frac{2\gamma_2}{\tau \omega_q^2 f_{12}}\right)^2}. \tag{12}$$

Interestingly, this expression implies that the performance of a dielectric polaritonic metasurface in absence of scattering or absorption losses does not depend on the actual doping density, but only on the quality factor of the intersubband transition $\gamma_2/\omega_q$, the relative scattering time $\tau\omega_q$ in the ionized electron gas, and the oscillator strength. For the present MQW, we have $f_{12} \approx 0.8$ and $\gamma_2/\omega_0 \approx 0.024$, and we estimate $\tau\omega_q = 22$. The reflection contrast for these parameters yields $|\Delta R| \approx 0.94$ or 94%. With the maximum oscillator strength of $f_{12} \approx 0.96$ the contrast increases to 0.95.

We can distinguish between local and nonlocal dielectric polaritonic metasurfaces. In the local approach, each element of the metasurface supports a Mie-like resonance. For example, strong coupling between an intersubband transition and a geometric resonance has been observed cylinders supporting a magnetic dipole [28]. Crucially, the magnetic dipole has a field that is largely z-oriented, enabling efficient interaction between the transition and the optical mode. Nonetheless, it should be noted that in Eq. (12) we assumed that all electric fields of the resonance are overlapping and aligned with the intersubband transition. In the dielectric metasurfaces, a fraction of the field will be stored outside of the dielectric quantum wells or not polarized in the vertical direction, and as a result $\Omega$ will be reduced, together with the resulting contrast.

Due to the small mode volume, local metasurfaces generally have a relatively low quality factor. Hence, to reach critical coupling and full absorption at saturation, significant absorption losses are also required, which in the present design are provided by the metal. In a full-dielectric design, absorption would be reduced significantly once the intersubband transitions are saturated, which means that achieving critical coupling to the small remaining losses is challenging with a local metasurface. This means that such local metasurfaces work best as saturable absorbers (e.g., as the current metasurface does at 41.5 THz in Fig. 3c), going from strong absorption at low intensity at one of the polaritonic branches to full reflection at saturation. One could also achieve limiting in local metasurfaces by using alternative "loss" mechanisms such as diffracted orders or transmission, but a more direct approach may be to use nonlocal dielectric metasurfaces instead. In these metasurfaces, the mode is delocalized and extends over many unit cells of a (quasi-)periodic structure. This enables much larger quality factors: in fact, such nonlocal metasurfaces may exhibit quasi-bound states in the continuum (Q-BIC) with arbitrarily high quality factors [36]. This control over the quality factor enables critical coupling to low absorption loss rates, which for example has already been utilized to achieve unity emissivity and absorptivity for thermal emitters [37]. Starting the design from a symmetry protected BIC with out-of-plane electric field ensures efficient coupling between the optical mode and the intersubband transition. Hence, depending on the desired application, using either local or nonlocal dielectric metasurfaces is a promising path towards even larger absorption contrasts than those demonstrated in the present paper.

## Discussion

In this work, we have theoretically and experimentally investigated optimized nonlinear metasurfaces based on ISB polaritons in MQW systems. We have provided a comprehensive theoretical analysis showing how the key metrics of these devices for limiting operations, such as the reflection contrast $\Delta R$, can be linked to the intrinsic material properties of the underlying MQW system and the metasurface parameters. Our analysis elucidates that the achievable values of $\Delta R$ can be largely increased by reducing the linewidth of the ISB transition ($\gamma_2$) and by minimizing the impact of higher-order electronic states (captured by the effective decay rate $\gamma_s$). In order to test these predictions, we optimized a MQW system based on $Ga_{0.53}In_{0.47}As/GaAs_{0.51}Sb_{0.49}$ heterostructures to produce ISB transitions with extremely narrow linewidths, more than a factor of 4 narrower than those used in previous works [19]. These new MQW materials allowed us to experimentally demonstrate nonlinear polaritonic metasurface limiters with absolute reflection contrast of about 54%, while simultaneously displaying record-high values of effective $\chi^{(3)}$. The measured values of reflection contrast are partially limited by the range of input powers available in our setup. Specifically, we estimate that if our measurements would include even lower input powers, a reflection contrast as high as 64% could be demonstrated within the same fabricated device. Additionally, we predict that, by using the same MQW material, the reflection contrast can approach values as high as 94% if the absorption induced by the metasurface could be removed – for example, by working with fully-dielectric designs.

Besides boosting the reflection contrast, the narrow linewidth of the ISB transitions also allowed us to observe fine-structure features in the metasurface absorption spectra that are induced by the bending of the semiconductor bands close to the metal contacts, resulting in additional ISB transitions with different transition frequencies. These additional states are crucial to correctly model the linear and nonlinear dynamics of polaritonic metasurfaces featuring giant nonlinearities. We anticipate that the properties of these additional ISB transitions may be tailored in future studies in order to further improve the metasurface performance as new degrees of freedom.

Overall, the results presented in this work provide an important step forward towards the design of efficient nonlinear polaritonic metasurfaces for application as saturable absorbers, power limiters, and nonlinear elements for all-optical control.

## Methods

*Experimental Setup*

The linear reflection spectra in Fig. 2 were acquired with a commercial FTIR spectrometer (Bruker INVENIO) coupled with a Hyperion microscope equipped with a low-NA ZnSe objective lens (Edmund, LFO-5-18). The spectra were normalized with respect to the reflection from the nearby gold substrate. The nonlinear reflection measurements (Fig. 3) were acquired with the custom-built setup shown in Fig. 5. The sample is excited with a pulsed laser (pulse duration $\tau$ = 2 ps, repetition rate f = 80 MHz) with central frequency tunable within the mid-IR. The mid-IR signal is obtained by difference-frequency generation module (APE, HarmoniXX DFG) which is fed by the signal and idler produced by an optical parametric oscillator (APE, Levante IR ps), which is in turn excited by a Yb pump laser (APE, Emerald Engine). A couple of holographic wire grid polarizers (LP1 and LP2, Thorlabs, WP25H-B) were used to control the power and polarization of the laser. The second polarizer (LP2) is set such that polarization of the laser is parallel to the long side of the antennas. The laser is mechanically chopped at 500 Hz, and then directed to a 50:50 ZnSe beamsplitter (Thorlabs, BSW710). Half of the signal impinges on a first HgCdTe detector (MCT1, Thorlabs, PDAVJ10). The other half of the signal is directed towards the sample. The signal is focused on the sample with a geltech aspheric lens ($L_{exc}$) with focal length f = 5.95 mm (Thorlabs, C028TME-F). The reflected signal is collected via the same lens and measured via a second (identical) HgCdTe detector (MCT2). Several neutral-density filters were used in front of both detectors to avoid saturation. Each detector is connected to a dedicated lock-in amplifier, and both lock-in amplifiers are referenced by the chopper controller. In order to measure the reflection level of the metasurface versus impinging power, the first polarizer (LP1) was systematically rotated in equal steps via a stepper motor (Thorlabs, K10CR1), and the corresponding signal generated by the two detectors was recorded. This procedure was repeated twice, first with the excitation beam focused on the metasurface and then with the beam focused on the bare substrate. By comparing the signal read by MCT2 in the two cases (and using the signal from MTC1 to monitor the laser power), the reflection level of the metasurface (with respect to the bare substrate) is then obtained. In order to measure the absolute power and intensity levels impinging on the sample, an additional calibration run was performed by measuring the average input power $P_{avg}^{(meas)}$ with a thermal powermeter (Thorlabs, S401C) placed before the lens $L_{exc}$. Due to the duty cycle of the chopper (50%), the actual input average power is $P_{avg} = 2 \times P_{avg}^{(meas)}$. Following standard formulas, the impinging peak intensity (both in time and space) was then calculated by

$$I_{peak} = \frac{2P_{avg}}{f\ \tau\ \pi w_0^2} \tag{13}$$

where $f$ = 80 MHz and $\tau$ = 2 ps. The spot radius $w_0$ (defined as the distance from the center of the spot at which the intensity drops by a factor $e^2$) was determined by both knife-edge measurements and by imaging the laser spot on a bolometric camera and comparing it with features of known sizes on the sample. We found that the spot radius is almost diffraction limited, and it increases linearly with the wavelength following the relation $w_0 \approx 0.73\lambda$.

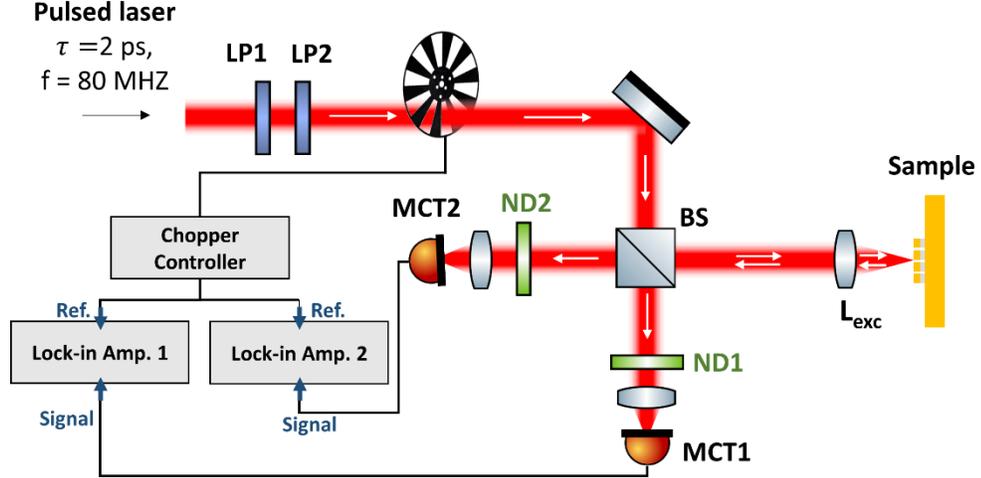

**Figure 5.** Experimental setup used to measure nonlinear reflection.

*Sample Design and Fabrication*

To fabricate the metasurfaces, a semiconductor heterostructure containing the MQWs and buffer layers is grown on top of a semi-insulating InP substrate using molecular beam epitaxy (MBE). The grown wafer is then coated with metal, thermocompressively bonded to the host wafer, stripped of the original InP substrate, and finally patterned into nanoantennas with the metal coating evaporated on top of the nanoantennas to form the double-metal resonators shown in Fig. 1(a).

The details of the heterostructure, including the substrate, are as follows, starting from the top layer of the original sample produced in the MBE:
1. Bottom GaInAs buffer layer, n-doping $1 \cdot 10^{19} \text{cm}^{-3}$, 5 nm thick
2. MQW structure, the structure is described in the main text, total thickness is 346 nm
3. Top GaInAs buffer layer (also serves as an etch-stop layer), n-doping $1 \cdot 10^{19} \text{cm}^{-3}$, 15 nm thick
4. 200 nm InP etch-stop layer, undoped
5. 1000 nm GaInAs etch-stop layer, undoped
6. 300 µm semi-insulating InP substrate

A metal layer composed of a titanium, platinum, and gold (layers thicknesses of 10 nm, 80 nm, 300 nm, respectively) is evaporated on top of the wafer. The wafer is then bonded episide-down to a similarly metal-coated n-doped InP host wafer (bonding temperature of 300°C, bonding pressure 1 bar). Subsequently, the substrate (layer 6) is removed using a concentrated HCl solution, followed by the selective removal of the etch-stop layers 5 and 4 using hydrogen peroxide-phosphoric acid solution mix and a diluted HCl solution, respectively. The 200µm x 200µm nanoantenna array patterns for each metasurface variant is lithographically defined using an electron beam system (Raith eLINE 30 kV) and a 100 nm-SiN hard mask. The MQW layer was then etched with Cl-Ar plasma in an inductively-coupled reactive ion etch (ICP RIE) system and the hard mask was removed using a $CF_4$-based RIE plasma. Finally, a 3 nm thick titanium and a 50 nm thick top gold layer was evaporated.

**Data availability.** Data underlying the results presented in this paper may be obtained from the authors upon reasonable request.

**Acknowledgements.** All authors acknowledge funding support from the DARPA Nascent Light-Matter Interactions program and the Air Force Office of Scientific Research MURI program. The TUM group acknowledges partial support from the German Research Foundation (Deutsche Forschungsgemeinschaft) grant number 506515587. The CUNY group acknowledges support from the Simons Foundation and the Air Force Research Laboratory.

**Competing interests.** The authors declare no conflicts of interest.

**Author Contributions.** J.K. designed and optimized the semiconductor heterostructure and fabricated the metasurface; M.C., S.M., J.K. performed the measurements; S.M. performed the modeling; S.M. and M.A.B. analyzed the ultimate limits; G.B. performed the semiconductor heterostructure growth; M.A.B. and A.A. supervised the work; all authors wrote the manuscript.

**Corresponding authors.** Correspondence to Andrea Alù, City University of New York, New York, NY 10016, USA, email: aalu@gc.cuny.edu; Mikhail Belkin, Walter Schottky Institute, Technical University of Munich, 85748 Garching, Germany, email: mikhail.belkin@wsi.tum.de.

**References**

[1] P. L. McMahon, *The Physics of Optical Computing*, Nature Reviews Physics 1 (2023).
[2] H. Baida, D. Mongin, D. Christofilos, G. Bachelier, A. Crut, P. Maioli, N. Del Fatti, and F. Vallée, *Ultrafast Nonlinear Optical Response of a Single Gold Nanorod near Its Surface Plasmon Resonance*, Physical Review Letters **107**, 057402 (2011).
[3] M. Ren, B. Jia, J.-Y. Ou, E. Plum, J. Zhang, K. F. MacDonald, A. E. Nikolaenko, J. Xu, M. Gu, and N. I. Zheludev, *Nanostructured Plasmonic Medium for Terahertz Bandwidth All-Optical Switching*, Advanced Materials **23**, 5540 (2011).
[4] Y. Yang, W. Wang, A. Boulesbaa, I. I. Kravchenko, D. P. Briggs, A. Puretzky, D. Geohegan, and J. Valentine, *Nonlinear Fano-Resonant Dielectric Metasurfaces*, Nano Letters **15**, 7388 (2015).
[5] M. R. Shcherbakov, P. P. Vabishchevich, A. S. Shorokhov, K. E. Chong, D.-Y. Choi, I. Staude, A. E. Miroshnichenko, D. N. Neshev, A. A. Fedyanin, and Y. S. Kivshar, *Ultrafast All-Optical Switching with Magnetic Resonances in Nonlinear Dielectric Nanostructures*, Nano Letters **15**, 6985 (2015).
[6] V. V. Zubyuk et al., *Low-Power Absorption Saturation in Semiconductor Metasurfaces*, Acs Photonics **6**, 2797 (2019).
[7] U. Keller, K. J. Weingarten, F. X. Kartner, D. Kopf, B. Braun, I. D. Jung, R. Fluck, C. Honninger, N. Matuschek, and J. A. Der Au, *Semiconductor Saturable Absorber Mirrors (SESAM's) for Femtosecond to Nanosecond Pulse Generation in Solid-State Lasers*, IEEE Journal of Selected Topics in Quantum Electronics **2**, 435 (1996).
[8] D. Miller, *Quantum-Well Self-Electro-Optic Effect Devices*, Opt. Quantum Electron **22**, S61 (1990).
[9] J.-L. Cheng, N. Vermeulen, and J. Sipe, *Third Order Optical Nonlinearity of Graphene*, New Journal of Physics **16**, 053014 (2014).
[10] X. Liu, Q. Guo, and J. Qiu, *Emerging Low-Dimensional Materials for Nonlinear Optics and Ultrafast Photonics*, Advanced Materials **29**, 1605886 (2017).
[11] L. Caspani et al., *Enhanced Nonlinear Refractive Index in ε-near-Zero Materials*, Physical Review Letters **116**, 233901 (2016).
[12] M. Z. Alam, I. De Leon, and R. W. Boyd, *Large Optical Nonlinearity of Indium Tin Oxide in Its Epsilon-near-Zero Region*, Science **352**, 795 (2016).
[13] M. Z. Alam, S. A. Schulz, J. Upham, I. De Leon, and R. W. Boyd, *Large Optical Nonlinearity of Nanoantennas Coupled to an Epsilon-near-Zero Material*, Nature Photonics **12**, 79 (2018).
[14] N. Kinsey and J. Khurgin, *Nonlinear Epsilon-near-Zero Materials Explained: Opinion*, Optical Materials Express **9**, 2793 (2019).
[15] Y. Yang, K. Kelley, E. Sachet, S. Campione, T. S. Luk, J.-P. Maria, M. B. Sinclair, and I. Brener, *Femtosecond Optical Polarization Switching Using a Cadmium Oxide-Based Perfect Absorber*, Nature Photonics **11**, 390 (2017).
[16] P. Vasa, W. Wang, R. Pomraenke, M. Lammers, M. Maiuri, C. Manzoni, G. Cerullo, and C. Lienau, *Real-Time Observation of Ultrafast Rabi Oscillations between Excitons and Plasmons in Metal Nanostructures with J-Aggregates*, Nature Photonics **7**, 128 (2013).
[17] N. T. Fofang, N. K. Grady, Z. Fan, A. O. Govorov, and N. J. Halas, *Plexciton Dynamics: Exciton-Plasmon Coupling in a J-Aggregate- Au Nanoshell Complex Provides a Mechanism for Nonlinearity*, Nano Letters **11**, 1556 (2011).
[18] A. Manjavacas, F. J. García de Abajo, and P. Nordlander, *Quantum Plexcitonics: Strongly Interacting Plasmons and Excitons*, Nano Letters **11**, 2318 (2011).
[19] S. A. Mann, N. Nookala, S. C. Johnson, M. Cotrufo, A. Mekawy, J. F. Klem, I. Brener, M. B. Raschke, A. Alù, and M. A. Belkin, *Ultrafast Optical Switching and Power Limiting in Intersubband Polaritonic Metasurfaces*, Optica **8**, 606 (2021).
[20] C. Ciuti, G. Bastard, and I. Carusotto, *Quantum Vacuum Properties of the Intersubband Cavity Polariton Field*, Physical Review B **72**, 115303 (2005).

[21] D. Dini, R. Köhler, A. Tredicucci, G. Biasiol, and L. Sorba, *Microcavity Polariton Splitting of Intersubband Transitions*, Physical Review Letters **90**, 116401 (2003).
[22] G. Günter et al., *Sub-Cycle Switch-on of Ultrastrong Light–Matter Interaction*, Nature **458**, 178 (2009).
[23] P. Jouy, A. Vasanelli, Y. Todorov, A. Delteil, G. Biasiol, L. Sorba, and C. Sirtori, *Transition from Strong to Ultrastrong Coupling Regime in Mid-Infrared Metal-Dielectric-Metal Cavities*, Applied Physics Letters **98**, (2011).
[24] M. Jeannin, E. Cosentino, S. Pirotta, M. Malerba, G. Biasiol, J.-M. Manceau, and R. Colombelli, *Low Intensity Saturation of an ISB Transition by a Mid-IR Quantum Cascade Laser*, Applied Physics Letters **122**, (2023).
[25] Y. Todorov, A. M. Andrews, I. Sagnes, R. Colombelli, P. Klang, G. Strasser, and C. Sirtori, *Strong Light-Matter Coupling in Subwavelength Metal-Dielectric Microcavities at Terahertz Frequencies*, Physical Review Letters **102**, 186402 (2009).
[26] E. Rosencher and P. Bois, *Model System for Optical Nonlinearities: Asymmetric Quantum Wells*, Physical Review B **44**, 11315 (1991).
[27] E. Benveniste, A. Vasanelli, A. Delteil, J. Devenson, R. Teissier, A. Baranov, A. M. Andrews, G. Strasser, I. Sagnes, and C. Sirtori, *Influence of the Material Parameters on Quantum Cascade Devices*, Applied Physics Letters **93**, (2008).
[28] R. Sarma et al., *Strong Coupling in All-Dielectric Intersubband Polaritonic Metasurfaces*, Nano Letters **21**, 367 (2020).
[29] R. Sarma, J. Xu, D. De Ceglia, L. Carletti, S. Campione, J. Klem, M. B. Sinclair, M. A. Belkin, and I. Brener, *An All-Dielectric Polaritonic Metasurface with a Giant Nonlinear Optical Response*, Nano Letters **22**, 896 (2022).
[30] D. M. Pozar, *Microwave Engineering* (John wiley & sons, 2011).
[31] S. Kohen, B. S. Williams, and Q. Hu, *Electromagnetic Modeling of Terahertz Quantum Cascade Laser Waveguides and Resonators*, Journal of Applied Physics **97**, (2005).
[32] T. Unuma, M. Yoshita, T. Noda, H. Sakaki, and H. Akiyama, *Intersubband Absorption Linewidth in GaAs Quantum Wells Due to Scattering by Interface Roughness, Phonons, Alloy Disorder, and Impurities*, Journal of Applied Physics **93**, 1586 (2003).
[33] C. Sirtori, F. Capasso, J. Faist, and S. Scandolo, *Nonparabolicity and a Sum Rule Associated with Bound-to-Bound and Bound-to-Continuum Intersubband Transitions in Quantum Wells*, Physical Review B **50**, 8663 (1994).
[34] H. Detz, A. M. Andrews, M. Nobile, P. Klang, E. Mujagić, G. Hesser, W. Schrenk, F. Schäffler, and G. Strasser, *Intersubband Optoelectronics in the InGaAs/GaAsSb Material System*, Journal of Vacuum Science & Technology B **28**, C3G19 (2010).
[35] P. Kordoš, M. Marso, R. Meyer, and H. Lüth, *Schottky Barrier Height Enhancement on N-In0. 53Ga0. 47As*, Journal of Applied Physics **72**, 2347 (1992).
[36] A. C. Overvig, S. C. Malek, and N. Yu, *Multifunctional Nonlocal Metasurfaces*, Physical Review Letters **125**, 017402 (2020).
[37] A. C. Overvig, S. A. Mann, and A. Alù, *Thermal Metasurfaces: Complete Emission Control by Combining Local and Nonlocal Light-Matter Interactions*, Physical Review X **11**, 021050 (2021).